\documentclass{emulateapj}
\slugcomment{DRAFT \today}
\shorttitle{Protostellar Radial Velocities}
\shortauthors{Covey et al.}

\begin{document}

\title{The Radial Velocity Distribution of Class I and Flat-Spectrum Protostars \footnote{The data presented herein were obtained at the W.M. Keck Observatory, which is operated as a scientific partnership among the California Institute of Technology, the University of California and the National Aeronautics and Space Administration. The Observatory was made possible by the generous financial support of the W.M. Keck Foundation.}}

\author{Kevin~R.~Covey\altaffilmark{2}, Thomas P. Greene\altaffilmark{3}, Greg W. Doppmann\altaffilmark{3,4}, Charles J. Lada\altaffilmark{5}}

\altaffiltext{2}{University of Washington, Department of Astronomy, Box 351580, Seattle, WA 98195}
\altaffiltext{3}{NASA Ames Research Center, Mail Stop 245-6, Moffett Field, CA 94035-1000}
\altaffiltext{4}{Present Address: Gemini Observatory, Southern Operations Center, Association of Universities for Research in Astronomy, Inc., Casilla 603, La Serena, Chile}
\altaffiltext{5}{Harvard-Smithsonian Center for Astrophysics, 60 Garden Street, Cambridge, MA 02138}


\begin{abstract}
We analyze radial velocities for a sample of 31 Class I and flat spectrum protostars in Taurus-Auriga, $\rho$ Ophiuchi and Serpens for evidence of the global dynamical state of extremely young stellar populations buried within parental molecular clouds.  Comparing the radial velocity of each protostar to that of the local CO gas, we are able to constrain the one dimensional radial velocity dispersion of Class I and flat spectrum objects to $\sim$2.5 km sec$^{-1}$ or below.  This upper limit to the protostellar velocity dispersion is consistent with the velocity dispersions of surrounding CO gas which we measure to be $\sim$ 1.4 km sec$^{-1}$, suggesting that the motions of protostars and local CO gas are dynamically linked and dominated by the gravitational potential of the molecular cloud.  However, the upper limit on the protostellar velocity dispersion could still allow for slightly inflated motions of protostars relative to the local molecular gas. Four of the protostars analyzed appear to have velocities more than 3 $\sigma$ (7.5 km sec$^{-1}$) away from the central local CO gas velocity while showing spectroscopic indicators of youth and accretion such as H$_{2}$ emission, HI Br $\gamma$ emission, or K band continuum veiling. These radial velocity outliers may represent protostellar spectroscopic binaries or ejected cluster members.  
\end{abstract}

\keywords{infrared:stars --- stars:formation --- stars:low-mass --- stars:pre--main-sequence --- stars: kinematics}

\section{Introduction}

The dynamical state of embedded protostellar populations is of crucial interest for the star formation process.  Until now, however, a diagnosis of the dynamical state of young (Class I and flat spectrum) protostellar populations has been largely inaccessible to direct observational study.  In this study we investigate the kinematics of low mass protostars and interpret the results within existing models of the star formation process.

Theoretical investigations have recently identified dynamical interactions between accreting protostars as a possible critical factor in the star formation process; this has been suggested both through analytic arguments \citep{Reipurth2001} and N-body simulations of star formation in a small cluster environment \citep{Bate2003}.  These studies indicate that gravitational encounters between protostars can separate a heavily accreting protostar from its mass reservoir, preventing further mass accretion and stunting its growth.  These interactions would predominantly affect the smallest members of an association, whose lower masses make them more susceptible to accelerations from gravitational interactions with nearby neighbors.  Accordingly, it has been suggested that dynamical interactions or ejections from an unstable multiple system may play a large role in shaping the mass spectrum in the substellar regime. 

Observational clues to the importance of the ejection mechanism in the context of the larger star formation process, however, are mixed.  Young brown dwarfs have been observed to possess infrared excesses indicative of circumstellar disks \citep{Wilking1999,Muench2001} and signatures of active accretion have been observed in their spectra as well \citep{Jayawardhana2003a,Luhman2003a}.  The predominance of steady mass accretion in isolated young brown dwarfs would seem to suggest any possible ejection mechanism must either not completely disrupt the accretion reservoirs of brown dwarfs (the very objects which should be most susceptible to disruption) or rare enough not to produce a sizable fraction of the brown dwarf population.  However, \citet{Muzerolle2003} find that analysis of $H\alpha$ emission from young brown dwarfs indicates an accretion rate of $\dot{M} \propto M^{-2}$ between 1 and 0.05 $M_{\odot}$, and that only 3 of 13 brown dwarfs in their sample appear to be strongly accreting.  These results may indicate that isolated brown dwarf accretion could be weaker than we might expect from linearly scaling T Tauri accretion to lower masses, possibly as a result of prior disturbances by interactions.  

A more direct probe of the ejection paradigm would be to directly measure the kinematic signature of ejection in sites of ongoing star formation.  We expect significant dynamical interactions between protostars will inflate the intrinsic velocity dispersion of the protostars above that of the local molecular gas, while a star formation process in which dynamical effects have little influence will result in a protostellar velocity dispersion similar to that of the local molecular gas from whence the protostars formed.  In short, an increase in the protostellar radial velocity dispersion above that of the local molecular gas serves as a proxy for the degree to which interactions affect the star formation process.  This intuition is buttressed by the results of numerical simulations of the star formation process, which show stellar interactions can increase the protostellar velocity dispersion by a factor of three above the velocity dispersion of the molecular cloud from which they formed \citep{Bate2003}.  Precise measurements, however, are required to test this model, as N-body simulations predict velocity dispersions even in such interacting environments at the 1-2 km sec$^{-1}$ level.

The dynamic state of protostellar populations is also interesting in the context of cluster formation and disruption.  The majority of young, forming stars in the Milky Way are to be found in cluster environments \citep{Lada2003}.  However, the stellar population of the Milky Way galaxy is dominated not by clusters but by the diffuse field population making up the thin and thick disk populations, presumably from stars which formed in now dissipated clusters \citep{Majewski1993}.  Explosive gas removal is thought to efficiently disrupt the majority of young clusters on timescales of 5-10 Myrs.  Those clusters which do survive the dispersal of their parental gas cloud eventually fall prey to evaporation and tidal heating, processes which operate on longer timescales of 10$^{8-9}$ years \citep{Lada2003}.  These effects are thought to leave measurable kinematic signatures on the populations they effect.  Mass loss from clusters, for example, is thought to result in a decreased velocity dispersion for the remaining bound members; the ratio of the initial and final velocity dispersions are correlated with the overall star formation efficiency in the cluster \citep{Mathieu1983}.  Measuring the dynamic state of protostellar populations can provide insight into mechanisms of cluster disruption by providing constraints on the initial velocity dispersion of newly formed stars for numerical simulations of cluster disruption, as well as by probing cluster evolution with a comparison of protostellar populations to more evolved bound clusters.  

The velocity dispersion of protostars in embedded clusters has not yet been observationally determined.  The radial velocity dispersion has been studied for more evolved T Tauri stars beginning in the work of \citet{Herbig1977} and continuing through the work of \citet{Jones1979}, \citet{Hartmann1986} and \citet{Dubath1996}.  These studies find no difference above the errors (typically $\sim$ 0.5 km/sec) between the mean stellar velocity and the central velocity of the local gas, and measure 1-D radial velocity dispersions of 2 km sec$^{-1}$ or less.  A study of this type has not yet been performed for protostellar populations, primarily because the spectroscopic resolution and sensitivity needed to measure radial velocities at the 2-3 km sec$^{-1}$ level has only recently become accessible in the infrared where these very young, highly embedded and reddened objects are easier to detect.  
 
Radial velocity measurements also serve to constrain the fraction of stellar systems which are members of close multiple systems.  Recent near infrared imaging surveys of Class I and flat spectrum sources in the Taurus, $\rho$ Ophiuchi, \& Serpens star formation regions \citep{Haisch2004, Duchene2004} show companion star fractions of $\sim 20-25 \%$ for separations between 300 and 2000 AU.  The companion star fraction of embedded protostars is thus significantly elevated above that detected for solar type main sequence stars in the same range of separations \citep[12\%;][]{Mathieu1994,Duquennoy1991a}, as also found for more evolved T Tauri stars in the same clouds. \citet{Duchene2004} also find that Class 0 and I sources with millimeter emission from protostellar envelopes have higher companion star fractions (38\%) than Class I/II/III objects without significant millimeter envelope emission (22\%).  These results suggest the multiplicity fraction of stars may decline through the pre-main sequence phase from high fractions in the most heavily embedded phases to lower fractions on the main sequence.   The observations analyzed here are sensitive to binaries with large orbital velocities, sampling binaries with much smaller separations than the studies of \citet{Haisch2004} and \citet{Duchene2004}, useful for probing possible effects of orbital evolution between the protostellar, T Tauri and main sequence phases.

In this study, we investigate the dynamical motions of Class I and flat spectrum protostars by studying the radial velocity dispersion of protostars about the local CO gas velocity.  In \S 2 we describe the observations used to derive radial velocities for Class I and flat spectrum protostars, as well as the observations which allow us to determine the systemic velocity of the CO gas associated with each star formation region.  Section 3 contains our analysis of the resultant radial velocity dispersion for Class I and flat spectrum sources, while \S 4 constrains the possible source of four prominent protostellar radial velocity outliers.  Section 5 discusses the implications of these results in the context of the standard picture of low mass star formation, and a summary of this work is contained in \S 6.

\section{Observations \& Velocity Determinations}

This study utilizes data from two distinct sources; the near IR spectral survey of Class I and flat spectrum protostars presented by \citet{Doppmann2005} (hereafter D05), and the CO survey of the Milky Way described by \citet{Dame2001}.  Though we refer the reader to the publications cited above for a full and complete discussion of the observations conducted by each survey, we provide below a brief summary of those observational details most relevant to the work presented here.  

\subsection{Protostellar Observations and Derived Radial Velocities}

The protostellar radial velocities used in this study were derived using sprectra from the Class I and flat spectrum survey conducted by D05.  Class I and flat spectrum sources were observed by D05 over the course of 12 nights between May 2000 and June 2003 using the near infrared echelle spectrograph NIRSPEC \citep{mclean1998} on the 10 m Keck II telescope.  Spectra were acquired with a 0.58'' (4 pixel) wide slit (allowing R $\sim$ 17,000), utilizing a 1024x1024 pixel InSb detector array with the NIRSPEC-7 blocking filter, under good ($\sim$ 0.6'') seeing.  The  NIRSPEC gratings were configured such that 4 prominent sets of spectral lines could be simultaneously observed in non-continuous orders: 1) the 2.1066 $\mu$m Mg and 2.1099 $\mu$m Al lines, 2) the 2.16609 $\mu$m HI $\gamma$ line, 3) the 2.2062 and 2.2090 $\mu$m Na doublet (blended with lines of Si and Sc) and the 2.2233 $\mu$m H$_{2}$ line, and 4) the 2.2935 $\mu$m CO bandhead.  We have selected for our analysis protostars with detected absorption features located in the $\rho$ Ophiuchus, Taurus-Auriga or Serpens star forming regions and observed by D05 in July 2001 or later, as the May 2000 data presented by D05 has much larger wavelength calibration uncertainties.  These criteria resulted in a sample of 31 protostars suitable for analysis in this work.  The set of MK spectral type standards observed by D05 have also been included in this analysis to provide estimates of observational uncertainties.  

A discussion of the methods used to reduce spectral data is presented in full by D05.  In summary, all spectra were flat-fielded, cleaned, sky subtracted, extracted, wavelength calibrated and corrected for telluric absorption using standard IRAF packages and procedures.  The radial velocity of each star was determined through cross correlation (using IRAFs FXCOR package) with spectral standards observed by D05 and with well measured radial velocities \citep{Nidever2002}.  The targets examined in this work range in spectral type from late G to mid M; given the onset of significant continuum structure for objects of types M0 and later, we divided the targets into pre and post M0 sub-samples for the purposes of cross-correlation.  Targets with spectral types earlier than M0 were cross correlated with template spectra of HD166620 (K2), HD201091 (K5) and HD219134 (K3).  Targets with spectral types of M0 or later were cross correlated with template spectra of Gliese 806 (M2), Gliese 4281 (M6.5) and VB8 (M7).  Cross correlation was performed on 3 echelle orders containing Mg, Al, Na and CO absorption features; each object was compared to all three applicable templates using all orders with detected absorption features.  A mean heliocentric velocity for each object was then found as the average of 9 independently derived velocity estimates (from 3 orders in each of 3 standards) in all but two cases.  Disk emission in EC38's CO order reduced the number of independent velocity estimates to 6, while WL12 produced extremely poor velocity estimates via cross correlation, so we have adopted D05's radial velocity estimate for this object.  The radial velocity of these protostars with respect to the local standard of rest (LSR) is presented in Table 1, after corrections for the solar motion with respect to the LSR were applied using IRAFs RVCORRECT task.  
 
 \begin{deluxetable}{lcccc}
\tablewidth{0pt}
\tablecaption{Radial Velocities of Class I and Flat Spectrum Protostars\label{tab:radv}}
\tablehead{
           \colhead{Target name} & 
	   \colhead{V$_{lsr}$} & 
	   \colhead{CO V$_{lsr}$} & 
	   \colhead{V$_{diff}$} &
	   \colhead{CO-stellar dist} \\ 
	   \colhead{} & 
	   \colhead{(km/sec)} & 
	   \colhead{(km/sec)} & 
	   \colhead{(km/sec)} &
	   \colhead{(beam widths)} }
\startdata

04016+2610  &  -10.42 & 6.83 &-17.25 & 0.56 \\
04489+3042  &  -18.27 & 6.83 &-25.10 & 0.61 \\
04295+2251  &    3.83 & 5.53 & -1.70 & 0.48 \\
Haro-6-28   &    3.64 & 5.53 & -1.89 & 0.26 \\
04108+2803B &    6.59 & 6.83 & -0.24 & 0.43 \\
DG-Tau      &    8.20 & 6.83 &  1.37 & 0.45 \\
GV-Tau S      &   -6.22 & 6.18 &-12.40 & 0.54 \\
04264+2433     &    6.72 & 6.18 & 0.54 & 0.40 \\
04158+2805  &    5.47 & 7.48 & -2.01 & 0.43 \\
04181+2655  &    7.55 & 6.83 & 0.72 & 0.39 \\
L1551-IRS5  &    7.89 & 6.83 & 1.06 & 0.55 \\
IRS43       &    3.01 & 3.74 & -0.73 & 0.43 \\
WL12        &    -0.61 & 3.48 & -4.09 & 0.19 \\
IRS67       &   3.26 & 4.78 & -1.52 & 0.78 \\
YLW16A      &    4.41 & 3.74 & 0.67 & 0.57 \\
162636      &    3.97 & 3.22 &  0.75 & 0.42 \\
GY21        &    2.65 & 3.22 & -0.57 & 0.83 \\
GY224       &    2.30 & 3.74 & -1.44 & 0.82 \\
VSSG17      &    7.06 & 3.48 & 3.58 & 0.68 \\
WL3         &   3.71 & 3.48 & 0.23 & 0.84 \\
WL17        &    4.88 & 3.48 & 1.40 & 0.89 \\
IRS51       &    6.30 & 3.74 &  2.56 & 0.12 \\
GY197       &    6.52 & 3.48 &  3.04 & 0.70 \\
GY91        &    2.91 & 3.48 &  -0.57 & 0.85 \\
CRBR12      &   -4.60 & 3.22 &-7.82 & 0.62 \\
EC125       &    5.07 & 7.91 & -2.84 & 0.92 \\
EC38        &   12.88 & 8.56 &  4.32 & 1.22 \\
EC91        &    5.66 & 7.91 & -2.25 & 0.77 \\
EC92        &    13.02 & 7.91 &  5.11 & 0.82 \\
EC94        &    9.70 & 7.91 & 1.79 & 0.79 \\
EC129       &    10.51 & 7.91 &  2.60 & 0.71 \\
EC53        &    13.36 & 8.56 & 4.80 & 1.26 \\

\enddata
\end{deluxetable}
 
To test the accuracy of the radial velocity determination resulting from this procedure, we have compared our derived heliocentric radial velocities with reported values for 15 spectral standards observed by D05 with accurately determined radial velocities \citep[errors $<$ 2 km sec$^{-1}$;][]{Mohanty2003, Nidever2002, Gizis2002,deMedeiros1999}.  Figure 1 displays the residuals obtained when the radial velocity reported in the literature is subtracted from our derived value.  We assume that the spread of non-zero residuals in Figure 1 is due primarily to observational error -- that is, we believe Figure 1 represents the expected observed radial velocity distribution for a population of stars with identical radial velocities ($\sigma$ = 0 km/sec) and observational errors typical of the limits of the observations presented by D05.  A best fit guassian to this distribution, shown in binned form in Figure 1 as a dashed histogram, results in a gaussian width ($\sigma$) of 1.28 km sec$^{-1}$.  We thus conservatively adopt 1.5 km sec$^{-1}$ as an estimate of the uncertainty in our velocity determinations for slowly rotating objects with strong absorption features.  

\begin{figure}[h]
\epsscale{1.2}
\plotone{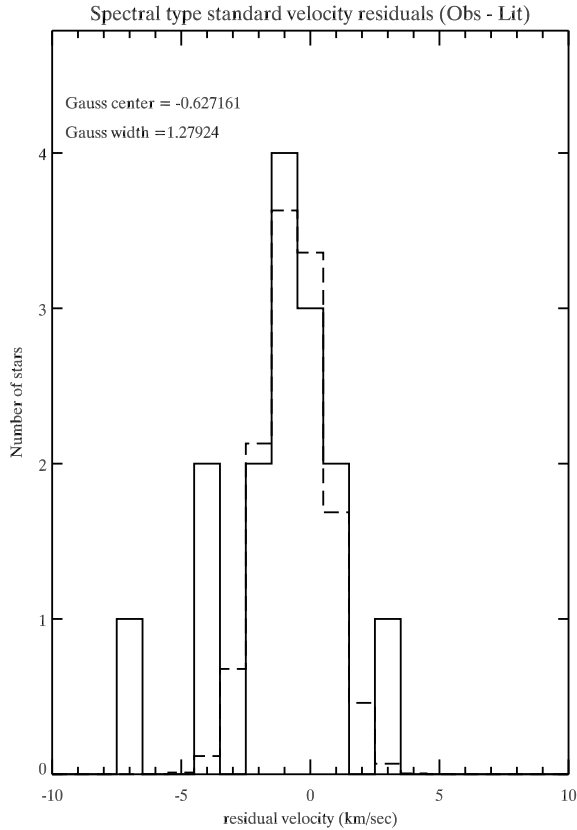} 
\caption {\scriptsize{Histogram display of residuals (solid line) when heliocentric velocities derived from NIRSPEC spectra of spectral type standards are compared to accurate (errors $\le$ 2 km sec$^{-1}$) determinations from the literature for the same stars.  The best fit guassian to the histogram, binned in the same manner, is shown as the dotted line.  The gaussian $\sigma$ provides an indication of the level of accuracy with which we can assign velocities ($\sigma \sim$ 1.3 km sec$^{-1}$).}}
\label{fig1}
\end {figure}

\subsection{Molecular Cloud Observations and Derived Radial Velocities}

The radial velocity dispersion of the protostars in each individual star forming region can be calculated directly from the V$_{LSR}$s reported in Table 1.  As we have only observed $\sim$ 10 protostars in each star formation region, individual radial velocity dispersions for each star formation region would be very sparsely sampled, possibly returning spurious results.  To improve the statistical leverage of our analysis, we have chosen to investigate the motions of protostars relative to the central velocity of the local molecular gas.  Assuming that the dynamical behaviors in the Taurus-Auriga, Serpens and $\rho$ Ophiuchi star formation regions are equivalent to first order, studying protostellar velocity residuals will allow us to combine results derived from protostars in different star formation regions, generating a snapshot of the dynamical state of newly formed stars.  The dispersion of protostellar radial velocities about the local gas velocity can also be compared to the velocity dispersion of the gas itself, allowing investigations of the extent to which stars and gas remain dynamically linked during the Class I/flat spectrum phase.  This enhanced statistical power comes at the expense of our ability to investigate differences in dynamics \textit{between} star formation regions.  Additionally, if ejection and evaporation are already important dynamical mechanisms by the Class I/flat spectrum stage, we might expect protostellar surveys to preferentially detect candidates whose motion has carried them to regions of the cloud with lower optical depths, as viewed from Earth.  This effect would be detected by a protostellar radial velocity distribution with a peak at lower (bluer) velocities than the associated CO gas, as we preferentially detect stars moving toward us and are less sensitive to objects moving away from us, deeper into the cloud.

\begin{figure}[h]
\epsscale{1.2}
\plotone{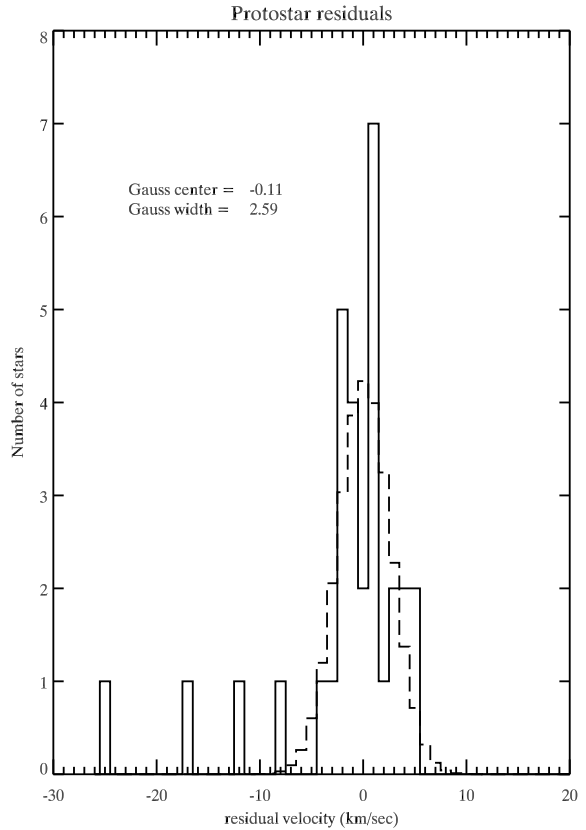} 
\caption {\scriptsize{Histogram display of protostar residual velocities (solid line; V$_{stellar}$ - V$_{CO}$).  The best fit gaussian to this distribution, binned in the same manner, is shown as a dashed histogram. The mean and $\sigma$ of the gaussian fit are displayed as well; though the protostars appear to have a wider velocity dispersion ($\sigma \sim$ 2.59 km sec$^{-1}$) than the standards shown in Figure 1, the rotational broadening and accretion veiling present in protostellar spectral features lead to more uncertain velocity determinations.  The expected magnitude of these effects match well the excess width of the protostellar velocity dispersion, suggesting we have yet to measure the true velocity dispersion of protostellar populations (see Section 3). }}
\label{fig2}
\end {figure}

To determine the difference between the radial velocity of each protostar and its local molecular gas, we relied on measurements from the CO survey of the Milky Way by \citet{Dame2001}. This survey homogeneously mapped all three star forming regions from which the protostars studied here were selected, sampled with a 0.125$^\circ$ beam (FWHM profile) and a grid spacing (measured in units of right ascension and declination) of 0.25$^\circ$ in the Serpens and $\rho$ Ophiuchus molecular clouds, and 0.125$^\circ$ spacing in Taurus. The central molecular gas velocity associated with each protostar in the sample was measured from the profile of the CO line at the nearest sampled position on the sky, usually within 1 beam-width of the protostellar position. CO linewidths were typically 1-2 km sec$^{-1}$. The central molecular gas velocity was found by bisecting the portion of the CO line which lies above $25\%$ of the peak CO emission at that location.  This method allows for a robust determination of the central cloud velocity which is insensitive to the precise line profile of the gas, a desirable quality given the optically thick and non-gaussian nature of many of the CO line profiles we examined.  Typical differences between line center velocities given by this method as opposed to standard gaussian fits to the CO profile are 0.2 km sec$^{-1}$, with maximum deviations up to 0.5 km sec$^{-1}$.  The central CO velocity at the nearest sampled position to each protostar is shown in the third column of Table 1, and the difference between the CO velocity and the protostellar velocity is shown in the fourth column.  The fifth column shows the distance between the protostellar position and the center of the CO sampling position, in units of the beam profile (FWHM).  

\section{The Protostellar Radial Velocity Dispersion}

Given the high degree of homogeneity in both the CO and NIRSPEC surveys, we are able to examine the radial velocity dispersion about the local CO systemic velocity of these protostars in bulk.  Shown in Figure 2 as a solid line is a histogram of the residual protostellar velocities about the central velocity of the local CO gas; a dashed line indicates the best fit gaussian to this distribution.  A comparison of the gaussian fits to the radial velocity distributions in Figures 1 and  2 show the protostellar population has a measured velocity dispersion almost twice that measured for our set of standards (standards $\sigma$ = 1.28 km sec$^{-1}$, protostellar $\sigma =$ 2.59 km sec$^{-1}$).  

The excess width of the protostellar velocity dispersion above that of the velocity dispersion of the radial velocity standards is likely due to factors which make the protostellar velocity determinations less reliable than those derived for the standards.  The first of these factors is the shallower and broader spectral features detected in protostellar spectra.  The presence of continuum veiling and moderate rotational velocities in these protostars result in spectral features which are shallower and broader than features in standards of similar spectral types.  The decreased strength of these features result in cross correlation functions whose central velocity is more uncertain, demonstrated by examining the standard deviation of velocities determined for the same object from different templates; radial velocities determined for standards from each of three appropriate templates show a median standard deviation of 0.8 km sec$^{-1}$, while the median standard deviation of the velocities derived for the protostellar sample is 1.2 km sec$^{-1}$, 50$\%$ larger.  Additional width may be introduced into the protostellar velocity dispersion given the uncertainty in the CO velocity to which each protostar is being compared -- while our observational uncertainty has been examined by comparing our derived velocities to known velocities with typical errors on the order of tens of meters per second, errors in the local CO velocity could be as large as 0.2 km sec$^{-1}$.  

The combination of weaker protostellar features and uncertainty in the local CO velocity implies the observational limits of the measured protostellar velocity dispersion are $\sim$ 0.6 km sec$^{-1}$ larger than of our standards, such at a zero velocity population would likely have a measured velocity dispersion of $\sim$ 1.9 km sec$^{-1}$.  This reduces the excess detected width of the protostellar velocity dispersion to 0.7 km sec$^{-1}$ above the level of expected observational uncertainties.  This excess width could be explained by an intrinsic protostellar velocity dispersion of 1.8 km sec$^{-1}$ added in quadrature to the estimated 1.9 km sec$^{-1}$ observational uncertainties.  It is equally plausible, however, that measuring the width of the velocity dispersion with a gaussian fit to a population with 31 members will allow additional statistical uncertainties due to sampling errors.

To test the statistical importance of any difference between the observed protostellar-CO velocity residuals and the observed radial velocity standard residuals, we have carried out a Kolmogorov-Smirnov (KS) test on the two sets of data.  The test is illustrated in Figure 3, which compares the cumulative probability functions of the protostellar and standard residuals.  The KS statistic measures the maximum difference between two cumulative distribution functions and can be used to calculate the probability that two distributions were drawn from the same parent sample \citep{Press1986}.  As seen in  Figure 3, the two distributions have a nearly 50\% chance of being derived from the same parent population.  

\begin{figure}[h]
\epsscale{1.2}
\plotone{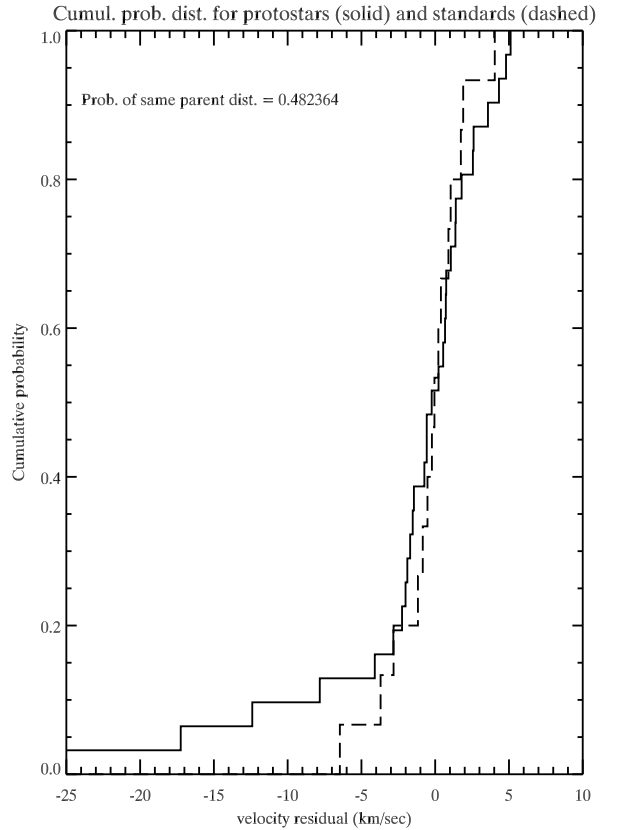} 
\caption {\scriptsize{Comparison between the cumulative probability functions of the standard and protostellar radial velocity residuals.  The probability the two functions originated from the same parent distribution, as estimated by an application of the KS test, is shown to be 48$\%$.}}
\label{fig3}
\end {figure}

We therefore find that the dispersion of protostellar radial velocities about the local CO gas velocity does not differ significantly from the same quantity as calculated for a sample of standards with accurately known radial velocities.  Given our observational errors, this constrains the width of the radial velocity dispersion of Class I/flat spectrum protostars to below $\sim$ 2.5 km sec$^{-1}$.  Within the limits of our measurements, therefore, we find no evidence that protostars and their local gas inhabit different dynamical environments -- that is, at the $\sim$ 2.5 km sec$^{-1}$ level, protostars and the local gas appear to be responding to the same gravitational potential.  Any effect of stellar interactions on the dynamical state of protostars must therefore lie below the $\sim$ 2.5 km sec$^{-1}$ level, or must be dominant only on timescales longer than that of the Class I/Flat Spectrum lifetime.

\section{Radial Velocity Outliers}

Although the radial velocity distribution of the protostars is consistent in bulk with that of the surrounding CO gas, we do detect four outliers (04016+2610, 04489+3042, GV Tau S \& CRBR 12) separated from the central velocity of the local gas by more than 7.5 km sec$^{-1}$.  Given our $\pm 2$ km sec$^{-1}$ radial velocity uncertainty, these objects represent $3 \sigma$ outliers from the central velocity of the local CO gas.  All four outliers possess negative (blueward) velocity shifts with respect to the local CO gas and other protostars in the same star forming region -- the outliers possess anomalous velocities, and are not outliers due to peculiar local CO velocities.  The three most significant outliers reside in the Taurus-Auriga star forming region.  These velocity outliers also appear to be bona fide protostars, and not older stars along the line of sight to the cluster: all 4 radial velocity outliers were found by D04 to have r$_{k} \ge$  0.9, all but CRBR 12 show HI Br $\gamma$ emission, and two (04016+2610, GV Tau S) show H$_{2}$ emission.  

\begin{deluxetable*}{lccccc}
\tablewidth{0pt}
\tablecaption{Maximum Orbital Periods for Possible Binaries\label{tab:bin}}
\tablehead{
           \colhead{Target name} & 
	   \colhead{Estimated M} & 
	   \colhead{Maximum Orbital Period} &
	   \colhead{Binary Separation} &
	   \colhead{Epoch of Obs.} \\ 
	   \colhead{} & 
	   \colhead{(M$_{\odot}$)} &
	   \colhead{(years)} &
	   \colhead{(AU)}&
           \colhead{(MM-DD-YYYY)} }
\startdata

04016+2610  & 1.6  &  2.08 & 2.40 & 11-04-2001\\
04489+3042  & 0.15  &  0.06 & 0.11 & 11-04-2001\\
GV-Tau S     & 1.6  &  5.55 & 4.62 & 11-06-2001\\
CRBR12      & 0.75  & 10.45 & 5.47 & 06-20-2003\\

\enddata
\end{deluxetable*}

There are 3 plausible explanations for such extreme velocity discrepancies in the radial velocities of young protostars: a) the protostars have had a close gravitational encounter with another member of the cluster and gained energy enough to be ejected from the cluster, b) the protostar is a member of a binary with a systemic velocity equal to that of the local CO gas, but was observed at a time when its primary's orbital velocity is directed in large part along our observational line of sight, or c) the velocities derived for these outliers have large uncertainties that are poorly characterized by our analysis of the velocities derived for stars with well known radial velocities.
 
The last two scenarios presented above can be easily confirmed or rejected with additional observations at later epochs.  It is plausible that the measurements reported here are significantly uncertain; three of these outliers have rotational velocities in excess of 25 km sec$^{-1}$, making cross correlation peaks more difficult to centroid, and three of these sources have been observed in the optical by \citet{White2004}, who find radial velocities for these objects (epoch Dec. 1999) that are consistent with other members of the Taurus star forming region.  For these reasons, additional observations to confirm the radial velocities reported here are necessary before any conclusions can be drawn as to the nature of these outliers.  However, to guide the interpretation of the resulting observations, we briefly examine below the possible ramifications of results of followup observations.

If multiple followup observations show in each epoch a protostar with a radial velocity equal to that of the local CO gas, it is safe to assume that these objects are outliers due to uncertainties in the analysis presented here (option c).  Additionally, if the radial velocities of these outliers are observed to change in a statistically significant fashion over the course of the follow up observations, it would be likely that their velocity excesses result from motions due to binarity (option b).  However, multiple followup observations showing radial velocities consistent with those reported here would not necessarily be proof that the stars in question are being ejected from the star forming region, as it is conceivable that the outliers could be members of a binary with a long orbital period.  

To constrain the timescales over which a consistent radial velocity in one of the outliers would need to observed before its possible binary status can be ruled out, we have constructed a simple model of each outlier as a binary system.  The binary parameters actually allowed for each source by the observed spectra depend on the specific T$_{eff}$, veiling, radius, and extinction (and spatial structure thereof) associated with each source.  In general, the observed veiling and lack of significant HI Br $\gamma$ absorption rule out the presence of hot, massive early type companions.  Accordingly, to provide upper limits on the orbital timescales these possible binaries should vary on, we will model each system as consisting of equal mass components -- generating the same radial velocity offsets with lower mass companions require smaller orbital separations and shorter orbital timescales.  

Assuming a canonical age for these outliers of 1 Myrs, the effective temperatures determined by D05 for the system can be translated into a mass via the models of \citet{Baraffe1998}.  Table 2 displays the estimated masses derived from the Baraffe models for each of these outliers -- in our simple model of these outliers as binary systems, we adopt the mass estimate given in Table 2 for each component of the binary.  The two hottest velocity outliers, GV Tau S and 04016+2610, have temperatures $\sim$ 250 K hotter than the hottest stars computed by Baraffe et al., requiring a small extrapolation to extend the grid to derive their likely masses.  

With mass estimates for the binary system components, we can now calculate the maximum timescale for changes in the observed radial velocity if the velocity separation from the CO velocity is due to a binary orbit centered on the mean CO gas velocity.  To calculate this maximum orbital period, we assume the observed systems consist of equal mass, single line spectroscopic binaries in circular, edge on orbits with orbital velocities equal to the velocity separation between the observed velocity and the local CO gas (shown in column 4 of Table 1).  Under these assumptions, the system's orbital period is related to the masses of the stars by:
\begin{equation}
\label{binary}
\frac{m_{comp}^3}{(m_{obs}+m_{comp})^2} = \frac{P}{2 \pi G} v_{obs}^3
\end{equation} 
where m$_{obs}$ and v$_{obs}$ indicate the estimated mass and detected residual velocity of the observed component, and m$_{comp}$ indicates the upper limit to the mass of any companion.  Our assumption of equal mass binaries implies that m$_{obs} = $ m$_{comp}$.  With the observed orbital velocity fixed, the assumptions of an equal mass binary in a circular, edge on orbit with maximal projection of the orbital velocity along the radial velocity direction drive the derived orbital period to the maximal end of the allowed range.  Relaxing those assumptions would predict shorter orbital timescales and more rapid radial velocity variations.  Thus, the timescales we derive for the completion of a full orbit represent upper limits -- changes in radial velocity may happen on shorter timescales, but should not occur on longer timescales unless the system has a primary with T$_{eff}$ greater than that detected by D05.  The resultant maximum orbital period for each outlier is shown in column 3 of Table 2, with corresponding binary separations given in column 4.  Typical orbital timescales for systems under these assumptions are on the order of several years, with typical binary separations of 2-5 AU.  

\clearpage

\section{Discussion}

The radial velocity measurements analyzed here represent the largest sample of radial velocity measurements for Class I/flat spectrum sources yet achieved.  These measurements allow investigation of a number of dynamical aspects of low mass star formation.

As shown in Section 3, the observed protostellar radial velocity dispersion ($\sigma$ = 2.59 km sec$^{-1}$) is not significantly broader than would be expected given the observational limits of this study.  These observations therefore place constraints on the width of the protostellar velocity dispersion to lie below the level of our observational uncertainties, or $\sim$ 2.5 km sec$^{-1}$.  Velocity dispersions of CO gas (as measured using gaussian fits) in these star formation regions are typically $\sim$ 1.4 km sec$^{-1}$; the upper limit placed on the protostellar velocity dispersion in this study is consistent with the local CO gas and protostars being dynamically linked during the Class I/flat spectrum stage, with little dynamical evolution from stellar interactions.  This upper limit does not, however, rule out possible dynamical effects which could inflate the protostellar velocity dispersion above that of the local CO gas, but below the 2.5 km sec$^{-1}$ level.  Future observations of protostars producing spectra with higher signal-to-noise ratios or more simultaneous wavelength coverage than obtained by D05 may allow for more accurately measured velocities via cross correlation with spectral templates, placing tighter constraints on the dynamical state of protostars.  Radial velocity uncertainties introduced by relatively large rotational velocities, however, may prove difficult to overcome.

Four of the 31 (12.9\%) Class I/flat spectrum sources appear to have radial velocities more than 3 $\sigma$ away from the local cloud velocity.  However, given these high $v$sin$i$s measured for these sources and their non-discrepant radial velocities as derived from optical spectra  \citep{White2004}, followup spectroscopy to confirm these radial velocities should be taken before these results are overinterpreted.  If real, these detections may present a lower limit to the close binary fraction of protostars.  \citet{Haisch2004} and \citet{Duchene2004} find companion star fractions of Class I/flat spectrum objects similar to that of more evolved T Tauri stars ($\sim 20-25\%$) in the star formation regions studied here.  The photometric nature of their study makes them most sensitive to companions with differences in K band apparent magnitude less than 4 to 6 magnitudes and at orbital distances from 300-2000 AU.  A radial velocity survey such as this is biased towards finding binaries at much closer radii: an equal mass binary system with 0.5 M$_{\odot}$ components in an edge-on, circular orbit with an orbital velocity of 7.5 km sec$^{-1}$ (the 3 $\sigma$ limit of this study) will have a separation of 3.9 AU and a period of 7.8 years.  Protostars with larger primary masses will have greater separations for similar orbital velocities.  Thus, the 12.9\% of stars detected in this work as radial velocity outliers could represent a single epoch lower limit to the number of protostars with close companions within $\sim$ 10 AU.   

The companion star fraction lower limit presented above, however, is uncorrected for incompleteness.  As part of a multi-epoch radial velocity survey for short period (P $<$ 100 days) T Tauri multiple systems, \citet{Melo2003} calculates incompleteness corrections for binaries with a 1.0 M$_{\odot}$ primary as a function of period (from 1 to 1000 days) and secondary mass.  Their typical detection efficiency is about 60\%, though dropping significantly towards longer periods.  Applying a similar correction to the number of radial velocity outliers seen here would imply a true companion star fraction of $\sim$ 22\%.  This would be roughly consistent with the companion star fraction measured for T Tauri stars over the same range of separations \citep[see Figure 6 of][interpolating over the gap in data for systems with periods between 100 and 10000 days ]{Melo2003}.  However, as this study is based on single epoch observations, the completeness fraction is likely significantly lower than 60\%, implying an even higher `true' companion star fraction, in excess of that observed for T Tauri stars at the same separations.  Given the good agreement between the T Tauri and protostellar companion star fractions at spatially resolvable separations, such a difference in multiplicity characteristics at close separations would be somewhat surprising.  As a result, it seems unlikely that all 4 radial velocity outliers are short period protostellar binaries.  

The four radial velocity outliers can also be interpreted as single stars with large intrinsic space velocities.  Such objects have presumably been ejected from their initial formation sites by dynamical interactions with other cluster members, as suggested for PV Ceph by \citet{Goodman2004}.  The blueward velocity shift of all 4 outliers could also be expected as a result of an ejection model, as introduced in Section 2.2.  However, the high velocities reported here (7.5 km sec$^{-1}$ and larger) are unusual even within the context of ejection models, as simulations of stellar encounters in a cluster environment \citep{Delgado-Donate2003,Bate2003} predict one dimensional velocity dispersions on the order of a few kilometers per second, and maximum ejection velocities of $\sim$ 5 km sec$^{-1}$.  The detection of spectroscopic accretion signatures also suggest the mass reservoir of these protostars have not been disrupted recently by dynamical interactions.  We therefore view interpretation of these high velocity outliers as a population of ejected cluster members as unlikely.   

However, our upper limit of $\sim$ 2.5 km sec$^{-1}$ on the one dimensional radial velocity dispersion of protostars is not inconsistent with predictions of models of star formation which include lower velocity dynamical effects.  As noted earlier, dynamical simulations of clustered star formation find 1-D radial velocity dispersions of 1-2 km sec$^{-1}$, comparable to the level of observational uncertainties in this study.  Additionally, halos of X-ray sources have been detected near sites of ongoing star formation in ROSAT X-ray surveys.  These halos typically extend to $\sim$ 5-10 degrees away from the center of the star formation region \citep{Feigelson1996,Neuhaeuser1997,Wichmann2000}.  The X-ray sources populating these halos possess Li absorption features, leading some to identify them as a dispersed population of Weak lined T Tauri Stars, while others find a lack of M type members of this population (which deplete their lithium on faster timescales than G \& K stars) indicative of an older population unassociated with ongoing star formation \citep{Briceno1997,Briceno1999}.  We find that the formation of dispersed halos of young X-ray sources could result naturally from the expansion of clusters whose radial velocity dispersions lie somewhat below the upper limits presented here and by \citet{Hartmann1986} for T Tauri stars.  For instance, at the distance of Taurus \citep[140 pc; ][]{Elias1978} a star moving 1 km sec$^{-1}$ in the plane of the sky can travel 7$^{\circ}$.5 in 18 Myrs, consistent with the size and estimated ages of X-ray halos.  Thus, a definitive analysis of the ability of pre-main sequence stars to produce diffuse halos of X-ray sources will require stronger constraints on the radial velocity distribution of star forming regions than are currently available.  

\section{Conclusions}

Our analysis of the radial velocities of Class I and flat spectrum protostars from three nearby star formation regions results in the following conclusions:

\begin{itemize}
\item{The one dimensional radial velocity dispersion of Class I/flat spectrum protostars is measured to be $\leq $2.5 km sec$^{-1}$.}
\item{The protostellar velocity dispersion upper limit presented here is consistent with the measured velocity dispersions of both CO gas and more evolved T Tauri stars in the same star formation regions.  This is consistent with the dynamics of the CO gas, protostars and T Tauri stars being dominated by the form of the large scale gravitational potential.  However, the 2.5 km sec$^{-1}$ protostellar radial velocity dispersion upper limit measured here is not in itself inconsistent with models of star formation in which dynamic stellar interactions play a significant role.}
\item{No large scale bias towards negative radial velocities is detected in this sample, as may be expected for stars with ejection velocities carrying them toward Earth and to areas of lower extinction.  The lower extinctions seen by these members would bias magnitude limited samples towards their inclusion, thus translating into a stellar median velocity offset from that of the local CO gas.}  
\item{We detect 4 of the 31 objects (12.9\%) studied here as 3 $\sigma$ outliers from the local CO velocity.  Further observations are needed to confirm these suprising velocities, which if true, imply either a large ($< 25\%$) close (r $<$ 10 AUs) binary fraction, or a significant level of dynamical ejections of young stars whose accretion reservoirs are nonetheless relatively undisturbed.  The 4 outliers detected here do have negative measured radial velocities, consistent with the blueward bias expected for ejected objects.}
\end{itemize}

\section{Acknowledgements}

The authors wish to recognize and acknowledge the very significant cultural role and reverence that the summit of Mauna Kea has always had within the indigenous Hawaiian community.  We are most fortunate to have the opportunity to conduct observations from this mountain.  The authors would like to thank George Herbig for an extremely rapid referee report; K.R.C is grateful to Anil Seth for useful discussions whose insights have improved this work.  All data have been reduced using IRAF; IRAF is distributed by the National Optical Astronomy Observatories, which are operated by the Association of Universities for Research in Astronomy, Inc., under cooperative agreement with the National Science Foundation.  This research has made use of NASA's Astrophysics Data System Bibliographic Services, the SIMBAD database, operated at CDS, Strasbourg, France, and the VizieR database of astronomical catalogues \citep{Ochsenbein2000}.  K.R.C gratefully acknowledges the NASA Graduate Student Researchers Program (NGT 2-52294) and T. Greene acknowledges NASA's Origins of Solar Systems Program for valuable support that enabled this work (Task 21-344-37-22-11).  


\end{document}